\documentstyle[epsfig,subfigure,a4,12pt]{article}
\epsfverbosetrue

\begin{document}
\hfill AS-TEXONO/00-06 \\
\hspace*{1cm} \hfill \today

\begin{center}
\large
{\bf
The Electronics and Data Acquisition Systems of \\
a CsI(Tl) Scintillating Crystal Detector for\\
Low Energy Neutrino Experiment\\
}
\normalsize
\vspace*{0.5cm}
W.P. Lai$^{a,b}$,
K.C.~Cheng$^a$,
H.B.~Li$^{a,c}$,
H.Y.~Sheng$^{a,d}$,
B.A.~Zhuang$^{a,d}$,
C.Y.~Chang$^e$,\\
C.P.~Chen$^a$,
Y.P.~Chen$^a$,
H.C.~Hsu$^a$,
J.~Li$^d$,
C.Y.~Liang$^f$,
Y.~Liu$^d$,
Z.S.~Liu$^g$,\\
C.S.~Luo$^a$,
F.~Shi$^d$,
R.F.~Su$^h$,
P.K.~Teng$^a$,
P.L.~Wang$^d$,
H.T. Wong$^{a,}$\footnote{Corresponding~author:
Email:~htwong@phys.sinica.edu.tw;
Tel:+886-2-2789-9682;
FAX:+886-2-2788-9828.},\\
Z.Y.~Zhang$^g$,
D.X.~Zhao$^{a,d}$,
J.W.~Zhao$^d$,
P.P.~Zhao$^d$,
Z.Y.~Zhou$^i$\\[2ex]
{\large
The TEXONO\footnote{Taiwan EXperiment On NeutrinO} Collaboration
}
\end{center}

\normalsize

\begin{flushleft}
$^a$ Institute of Physics, Academia Sinica, Taipei, Taiwan.\\
$^b$ Department of Management Information Systems,
Chung Kuo Institute of Technology, \\
\hspace*{1cm} Taipei, Taiwan.\\
$^c$ Department of Physics, National Taiwan University, 
Taipei, Taiwan.\\
$^d$ Institute of High Energy Physics, Beijing, China.\\
$^e$ Department of Physics, University of Maryland, College Park, U.S.A.\\
$^f$ Center of Material Science, National Tsing Hua University, 
Hsinchu, Taiwan.\\
$^g$ Department of Electronics,
Institute of Radiation Protection, Taiyuan, China.\\
$^h$ Nuclear Engineering Division, Kuo-sheng
Nuclear Power Station, \\
\hspace*{1cm} Taiwan Power Company, Taiwan.\\
$^i$ Department of Nuclear Physics,
Institute of Atomic Energy, Beijing, China.\\
\end{flushleft}

\newpage

\begin{center}
{\bf
Abstract
}
\end{center}

A 500 kg CsI(Tl) scintillating crystal detector is under
construction for the studies of low-energy neutrino physics. 
The requirements, design, realization
and the performance of the associated 
electronics, trigger, data acquisition and software control systems 
are described. Possibilities for future extensions are discussed. 

\vspace*{0.1cm}

\begin{flushleft}
\end{flushleft}

\begin{center}
{\it Submitted to Nucl. Instrum. Methods.}
\end{center}

\clearpage

\section{Introduction}

\begin{figure}
\centerline{
\epsfig{file=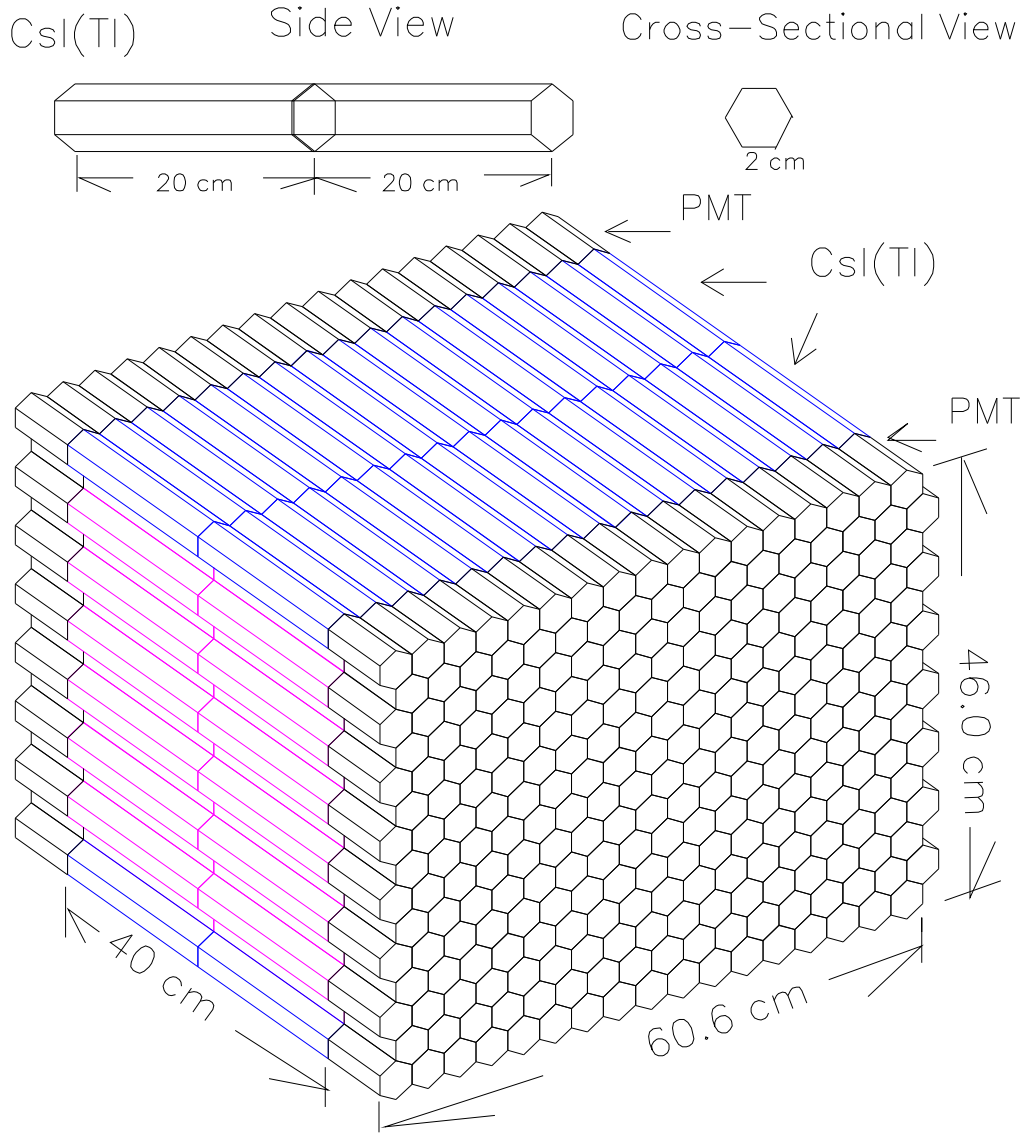,width=15cm}
}
\caption{
Schematic drawings of the CsI(Tl)
target configuration, showing a 2(Width)X
17(Depth)X15(Height) matrix.
Individual crystal module is 20~cm long with
a hexagonal cross-section of 2~cm edge. Readout
is performed by photo-multipliers
at both ends.
}
\label{fig:setup}
\end{figure}

One of the major directions and experimental
challenges in neutrino physics~\cite{bib:nuphys}
is to extend the measurement capabilities to
the sub-MeV range for the detection of
the p-p and $^7$Be solar neutrinos and other
topics.
The merits of scintillating
crystal detector in low-energy low-background
experiment were recently discussed~\cite{bib:prospects}.
An experiment with a CsI(Tl) detector
placed near a reactor core to study
neutrino interactions at low energy is
being constructed~\cite{bib:expt}.
In the first data-taking phase,
the detector is based on 
200~kg of CsI(Tl) scintillating
crystals\footnote{Manufacturer: Unique Crystals, Beijing}
read out by 200 photo-multipliers (PMTs)\footnote{Hamamatsu CR110-10}
made of special low-activity glass.
The set-up is shown schematically in
Figure~\ref{fig:setup}.
The eventual goal will be to operate
with 500~kg of CsI(Tl) crystals.

The tasks and requirements of the
readout systems are to
achieve a low detection threshold,
large dynamic range and
good energy resolution provided by
the CsI(Tl) and PMTs,
to provide relative timing among individual
channels as well as delayed correlated events
for background diagnostics, and 
to record the pulse shape faithfully
for pulse shape discrimination between
$\gamma$/e and $\alpha$-particles~\cite{bib:prototype}. 
The sum of the two PMT signals for each
crystal module gives the energy while
their difference provides the longitudinal
position information.
The intrinsic rise and decay times  
for the emissions from CsI(Tl) 
are typically at the range of 100~ns and
$\rm{2~\mu s}$, respectively.
A typical signal, as recorded by
a 100~MHz digital oscilloscope, 
is shown in Figure~\ref{fig:signals}a.

This article describes
the requirements, design, construction 
and performance of the
the electronics, data acquisition (DAQ),
and software control systems
of this experiment.

\section{Electronics System}
\label{sect::ele}

\begin{figure}
\centerline{
\epsfig{file=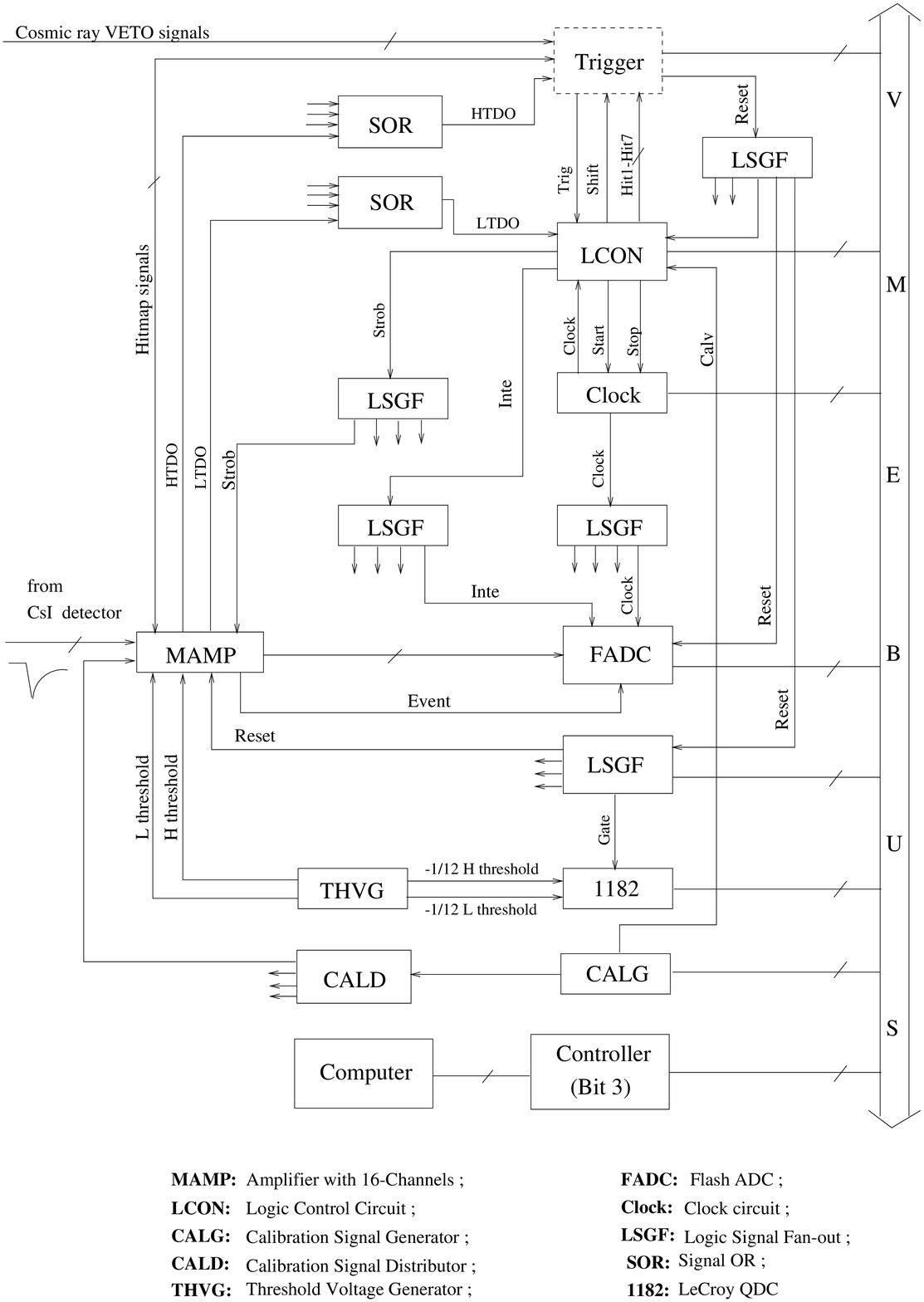,height=18cm,width=1.0\textwidth}
}
\caption[Electronics]{
A schematic diagram of the electronics system.}
\label{fig:Electronics} 
\end{figure}

The schematic diagram of the electronics system is
depicted in Figure~\ref{fig:Electronics}.
Every channel of the ``raw''
`(CsI(Tl)+PMT) signal 
is amplified and shaped by the
amplifier/shaper (AS), and subsequently
digitized by the flash analog-to-digital convertor (FADC).
The trigger system  selects relevant events
to be read out, 
while the 
logic control system provides
a coherent timing and synchronization
for the different electronics modules.   
The stability and linearity of the system
is monitored (and corrected for, if necessary)
by the calibration unit.
For optimal adaptations to our applications,
as well as for cost-effectiveness reasons,
almost the entire electronics system is
designed, constructed and tested by the
Collaboration.

Two industry standards are adopted: NIM-convention
for the AS and the various coincidence and veto
logic modules,
while the VME-protocols  for the 
electronics-data~acquisition interface.
The trigger modules, logic control unit and calibration pulser   
are ``double height'' VME-6U modules, while 
the FADCs are ``triple height'' 
VME-9U modules.
All are single-width modules.

\begin{figure}
{\bf (a)}
\centerline{
\epsfig{file=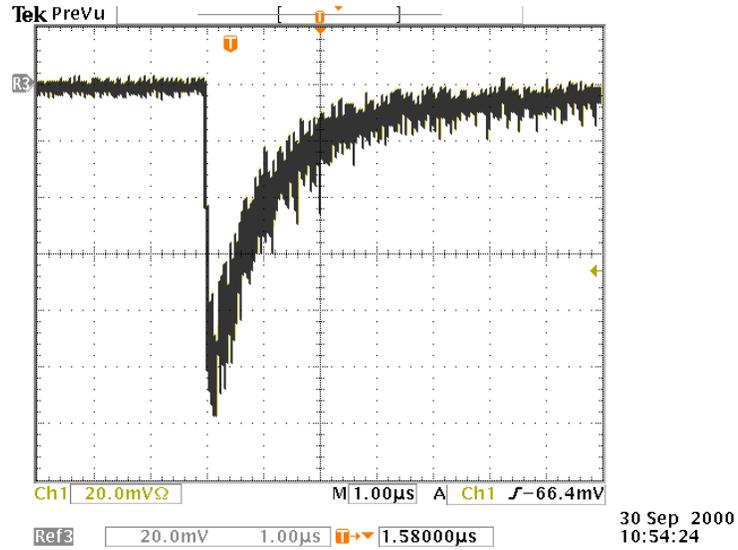,width=10cm} 
}
{\bf (b)}
\centerline{
\epsfig{file=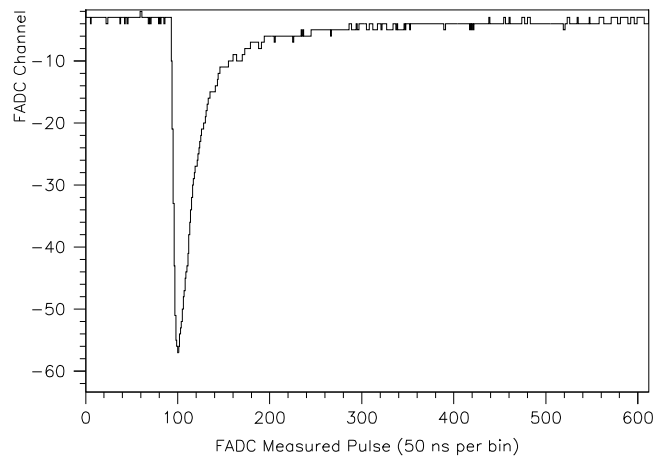,width=10cm}
}
\caption{
(a) Raw input signal from CsI(Tl)+PMT as recorded by
a 100~MHz digital oscilloscope. 
Time axis: 1~$\mu$s per box.
(b) Output signal
after shaping from the Amplifier/Shaper as recorded
by the FADC.
Time axis: 5~$\mu$s per 100~FADC time bin.
}
\label{fig:signals}
\end{figure}

The PMT signals pass through
discriminators embedded in the AS module,
and the NIM-level output are 
transferred to the trigger system. 
Simultaneously, the 
pre-trigger pedestals and the 
amplified  AS pulses are continuously digitized 
and recorded on the circular buffers of 
the FADCs running at a 20~MHz (50~ns) rate. 
A typical FADC output signal of
an event originated from the CsI(Tl) crystal
is depicted in Figure~\ref{fig:signals}b.
The shaping effects can be seen when compared
to the raw CsI(Tl)+PMT pulse in Figure~\ref{fig:signals}a.
The integrated area of the signal from two PMTs added together
gives the energy of the event.
  
Once a signal due to a valid trigger is 
issued,  
the logic control system will 
generate a pre-defined ``digitization gate''
which determines the FADC digitalization
time duration for the PMT signals. 
At the end of the gate, the FADC recording stops,
and an ``Interrupt'' request is issued to the DAQ software hosted 
on a Linux-based PC via the VME-bus and a 
commercial VME-PCI bus bridge controller. 
The entire FADC system issues 
a single DAQ interrupt request (instead of each
per module).
The DAQ software response accordingly,
and the data from the FADC circular buffers are
read out  and saved at
computer disks 
for further processing.  A detailed
description of the online/offline DAQ system is given in 
Section~\ref{sect::software}. 

The calibration pulser generates and distributes
simulation signals to monitor the
functionalities and performances of all channels, 
helps trouble-shooting for locating noisy
or dead channels and provides them with
correction factors for subsequent analysis.

The components of the electronics system are discussed in
the following sub-sections.

\subsection{Amplifier and Shaper}

\begin{figure}
\epsfig{file=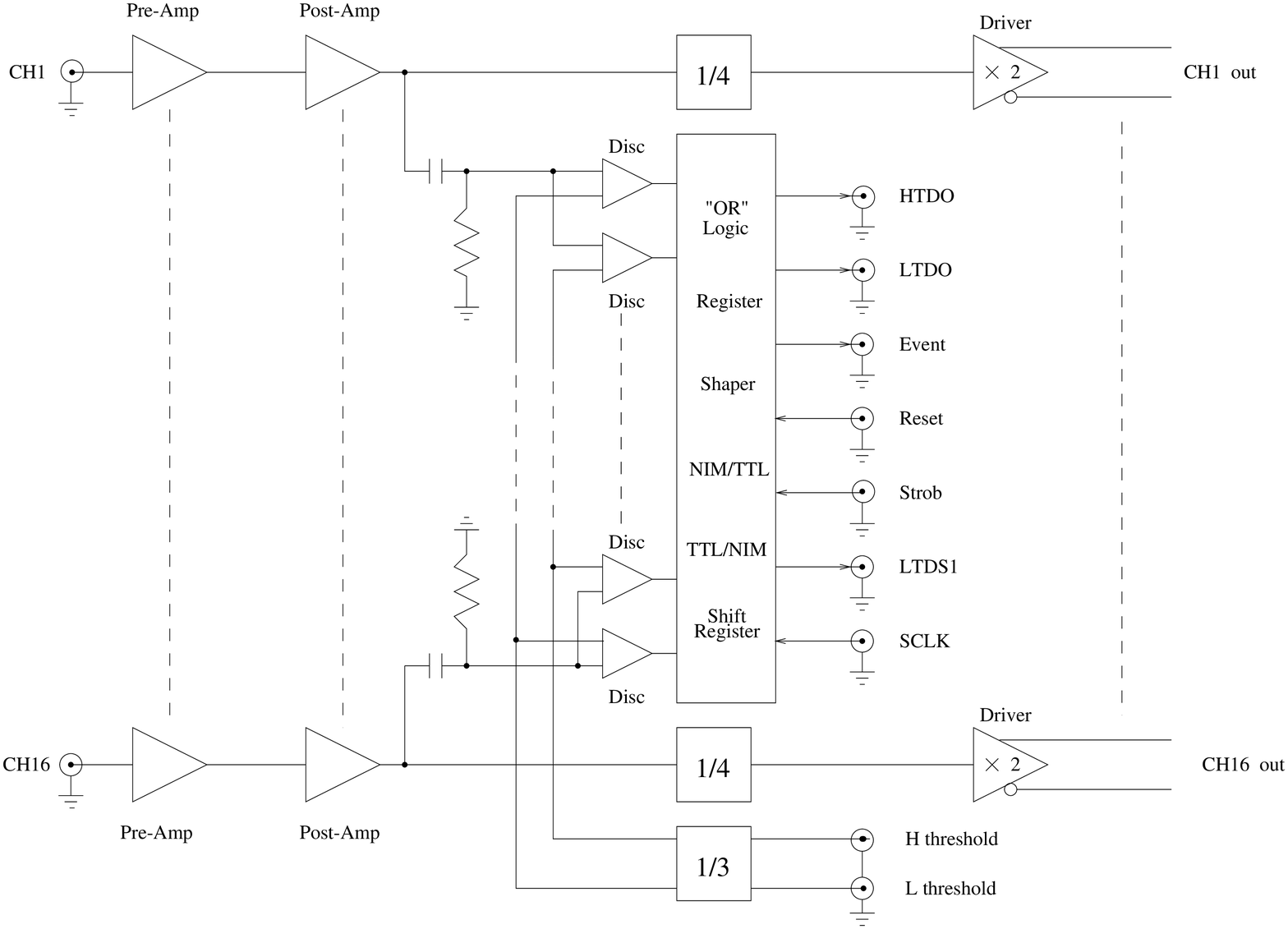,height=18cm,width=1.0\textwidth}
\vspace{0.5cm}
\caption[MAMP]{\label{fig:MAMP} 
A schematic diagram of the amplifier/shaper module, 
which contains 16 channels each with 2 discriminator
outputs. }
\end{figure}

The schematic diagram of the 
amplifier/shaper (AS) module is
shown in Figure~\ref{fig:MAMP}.
The AS receives 
the PMT signals via LEMO cables
and provides them with amplification, filtering,
and pulse shaping.
In addition, two 
discriminators per channels are integrated in the same
module.
A total of 16 channels are grouped 
into one module for the AS and the FADC. 

The discriminator threshold levels
for the PMT signals are
provided by a 
separate manually-set threshold voltage adjustor
and distributor module.
The high one (``HiThr'') is for
trigger purposes while the low one (``LoThr'')
allows recording of all signals
minimally above electronics noise.
The settings are constantly monitored by a
commercial VME Charge-to-Digital Convertor (QDC)
board\footnote{\label{qdc} LeCroy ADC 1182}.
The discriminator signals, ``HDTO'' and ``LDTO''
for those passing HiThr and LoThr, respectively, 
are sent to the trigger for processing
by the trigger system. 


As shown in Figure~\ref{fig:signals}a,
the raw signals originate
as a summation of a collection
of single photo-electrons (pe) 
statistically distributed over
a time duration
according to the standard light curve
of the CsI(Tl).
Therefore, they are 
several $\mu$s in length whose profile 
are rough with O(10~ns) spikes 
due to single pe.
In comparison, thermal noise from
PMT appear as the single pe 
spike with this O(10~ns) width.

The first stage\footnote{Based on Harris IC HA-2525}
of the AS
provides shaping at a time constant of 27~ns.
It functions as a charge-sensitive amplifier
for short [O(10~ns and less)] signals while
remains as a current-sensitive amplifier
for long ($>1~\mu$s) signals.
As a result,
the single pe spikes
are smoothened
relative to the digitization
time-bin of 50~ns,
such that their amplitude are 
reduced to below the LoThr level.
Consequently,
the trigger threshold
can be reduced, and that the 20~MHz 
FADC clock rate are adequate and appropriate.
The subsequent 
stage\footnote{Based on Analog Devices IC AD817}
is a slow device
providing effective shaping 
of about 250~ns and 1.6~$\mu$s in rise
and fall times, respectively.
The overall gain is 3.2~V/mA.

Each AS module is also equipped 
with a 16-bit shift register, which 
can give a hit pattern of the 16 channels on the module. 
The hit-pattern can be used to generate trigger decisions
and to 
reduce the reading time of the FADC module.
However, in a low count-rate experiment like
ours, the DAQ dead time is not a of critical
concern, and therefore, this
functionality of fast hit-pattern measurement
is not implemented for the first operational
version of the electronics system.

\subsection{Flash ADC}

\begin{figure}
\epsfig{file=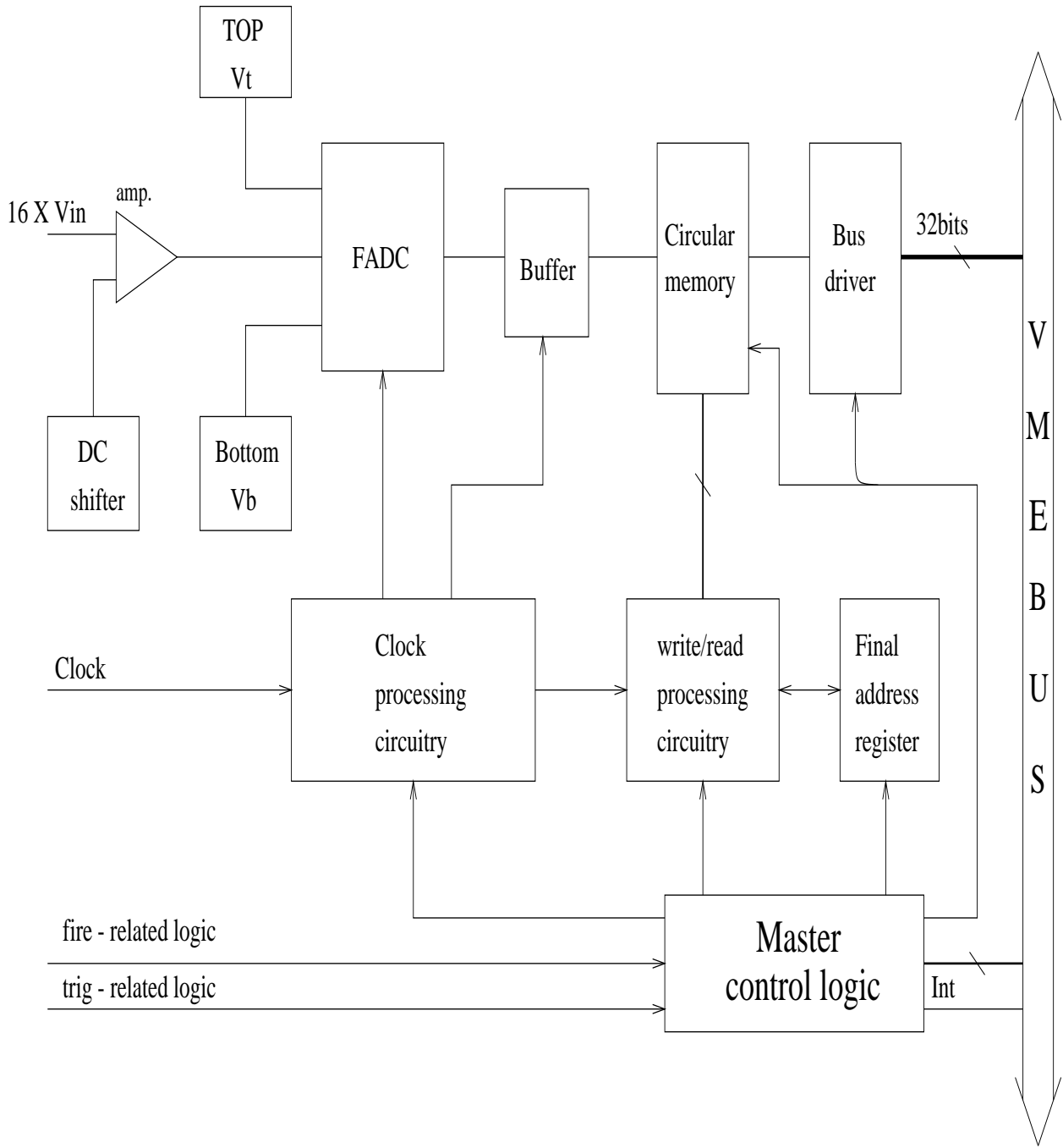,height=18cm,width=1.0\textwidth}
\vspace{0.5cm}
\caption[FADCmodule]{\label{fig:FADCmodule} 
A schematic diagram of the FADC circuitry.
There are 16 channels in a module.}
\end{figure}

The main tasks of the FADC are to digitize the pulse 
from the main amplifier/shaper, 
and to issue an Interrupt request  to the DAQ system. 
The schematic diagram of the FADC is displayed in
Figure~\ref{fig:FADCmodule}.
There are 16 channels per module, matching
the differential output signals from the AS 
delivered to the FADC with flat twisted cables.
The resolution is 8-bit and
the gain is adjusted such that the full
dynamic range corresponds to an input of
2~V.

After an event is read out and the DAQ system is
ready to receive the next event, a ``Reset'' signal
is issued to the electronics hardware.
The input signal (which are the AS output) 
are continuously 
digitized\footnote{Based on Analog Devices IC AD775},
and the data are  put into 
a circular buffer memory\footnote{Elite MT IC UM61256FK}.
of size 4096~bytes (4K in depth and each
bin having 8~bit resolution).
The digitization frequency is driven by 
an externally programmable clock. The
hardware limiting rate is 30~MHz, while
typically a 20~MHz clock is used.

Digitization is terminated when the
external clock stops at the end of a valid event:
that is, typically 25~$\mu$s after a trigger is
issued at ``t=0''. Any subsequent events up  
to a pre-selected time (typically 2~ms) 
are also recorded.
The circular 
buffer points at the last byte it writes.
The valid data on the circular 
buffer is then read out in the time-reversed sense,
starting with the last byte until some pre-trigger 
time bins (typically 5~$\mu$s before t=0)
which are for extracting the
pedestal information.

To enhance the reading speed, the 16 channels on one module are 
divided into four groups, and each group takes a 32-bit word for data
transfer since the circular buffer for each channel uses 8-bit 
per recording. 
Based on special features to be described
in details in Section~\ref{sect::logic},
only those FADC modules with at least one
hit above LoThr within the event duration
will be read out.

\subsection{Calibration Pulser}


The calibration pulser provides the information 
of dynamical range, linearity, 
and stability for all channels. 
It helps trouble-shooting
to locate problematic channels,
and provides
a set of online or offline correction 
parameters.

\begin{figure}
\epsfig{file=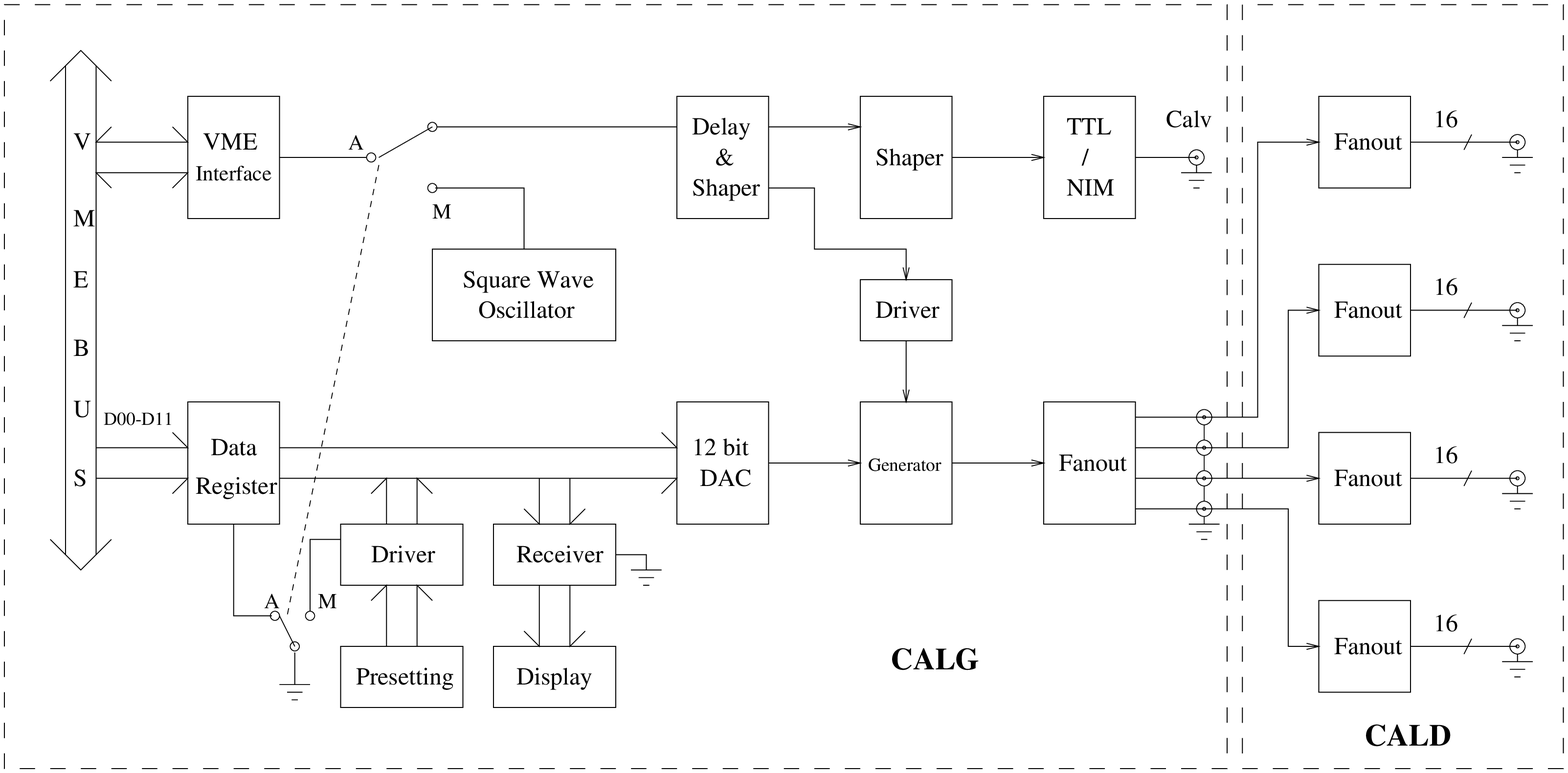,height=18cm,width=1.0\textwidth}
\vspace{0.5cm}
\caption[Calpulser]{\label{fig:Calpulser} 
A schematic diagram of the calibration pulser circuit. }
\end{figure}

As illustrated by
the schematic diagram 
in Figure~\ref{fig:Calpulser},
the pulser module
consists of two components: 
calibration signal generator and distributor. 
The former generates a simulated signal, 
which is then fanned out via the latter to provide
the input of the AS modules
for all channels. 
A non-linearity of better than $0.3\%$ 
has been achieved for all channels. 

During steady-state data taking, the
calibration will be performed several times in
a day to the electronics
system via fully automatic software.

\subsection{Logic Control Unit}
\label{sect::logic}

\begin{figure}
\epsfig{file=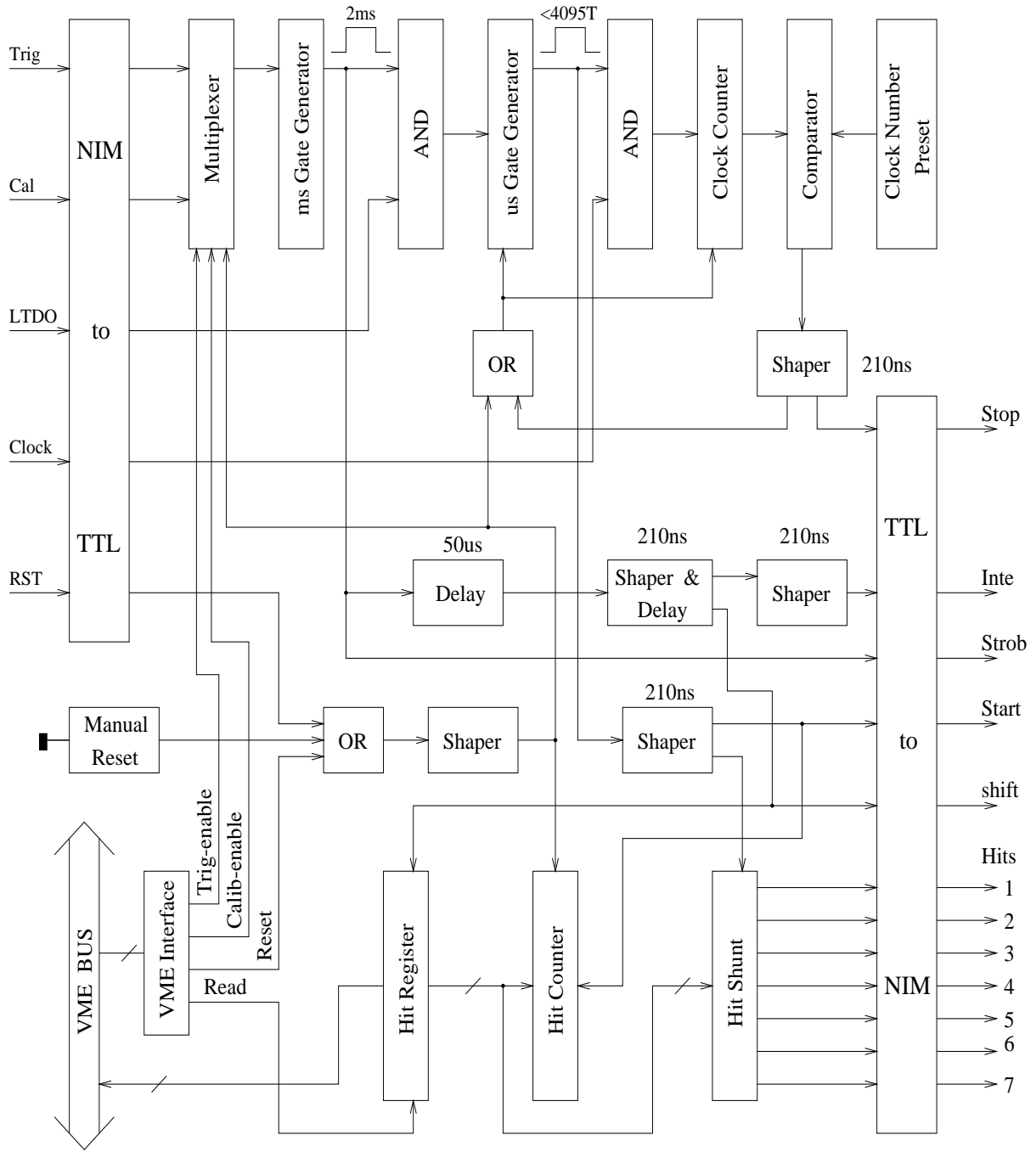,height=18cm,width=1.0\textwidth}
\vspace{0.5cm}
\caption[LogicControl]{\label{fig:LogicControl} 
A schematic diagram of the logic control unit.}
\end{figure}
\begin{figure}
\epsfig{file=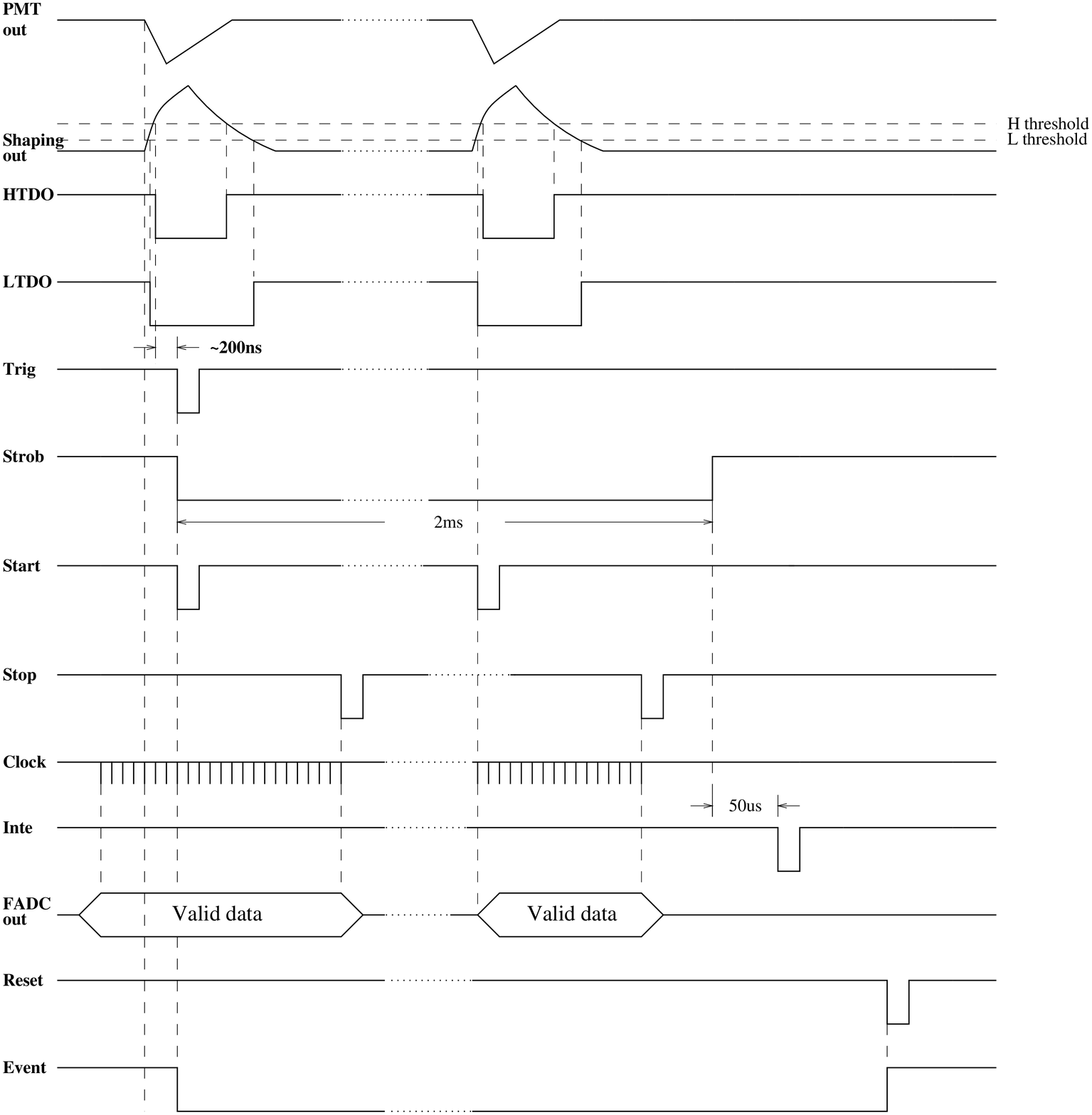,height=18cm,width=1.0\textwidth}
\vspace{0.5cm}
\caption[Timing]{\label{fig:Timing} 
The timing sequence in a typical event.}
\end{figure}

The schematic diagram of the logic control
unit is depicted in
Figure~\ref{fig:LogicControl}. 
Its main function is to coordinate
and synchronize the interactive actions 
of the various components of
the electronics and DAQ systems.
The designed timing schematic is shown in
Figure~\ref{fig:Timing}. 

After receiving a valid trigger 
signal which defines the ``START'' timing, 
control signals
are distributed by
the logic control unit  
to the various components.
The hardware pre-selectable
digitization gate, typically of 25~$\mu$s width,
is initiated.
At the end of the
gate,  a  ``STOP'' signal is generated
which disables the clocking circuit and 
the FADC digitization will stop.

For detailed diagnostics of the background
events,  it is desirable to record the
possible occurrence of delayed correlated
events, such as
the $^{214}$Bi-$^{214}$Po decay sequence
\begin{displaymath}
\rm{
^{214}Bi ~ \rightarrow ~ ^{214}Po ~ + ~ \bar{\nu_e} ~ +
~ e^- ~ + ~ \gamma 's ~  (Q=3.28~MeV ~ ; ~ T_{\frac{1}{2}}=19.8~min) ~,
}
\end{displaymath}
\begin{displaymath}
\rm{
^{214}Po ~ \rightarrow ~ ^{210}Pb~ + ~ \alpha ~
(Q=7.69~MeV ~ ; ~ T_{\frac{1}{2}}=164~ \mu s) ~~~.
}
\end{displaymath}
Full digitization of the entire period 
far exceeds the available memory space in the
circular buffer. 
To realize this requirement,
a special ``cascade-sequence''   
function  is implemented
After the completion of the 
triggered event, the entire system
remains at the read-enable
state capable of recording
the pulse shape as well as timing of
any delayed activities above LoThr for
the duration of the ``STROB'' gate, 
typically set at 2~ms.
Delayed events will initiate another
cycle of digitization gate during which
FADC digitization resumes, starting
from the last location of the previous
events.
Relative arrival times of these
delayed events are measured by
additional Time-to-Digital-Convertor (TDCs)
with 1~$\mu$s resolution.
In this way, complete information
up to 7~delayed events
can be recorded (for a 20~MHz clock rate
at the memory depth of 4K). 
In addition,
the computer clock time is 
read out for every event, providing
complementary timing information between
events. 
The measured timing uncertainty is close
to the limiting accuracy from
the computer clock resolution of 10~ms.
Therefore, the detailed and complete
timing sequence of all the events
can be reconstructed offline,
providing powerful diagnostic tools for
background understanding and suppression.

A manual switch provides selection
of data taking modes either with the 
cascade-sequence functionalities
or in 
the single-event mode for
calibration purposes.

An Interrupt (``INTE'') signal is
issued to the DAQ system at the
completion of the STROB gate.
The DAQ system starts transferring
the data from the FADC memory and
other locations to the hard disk
storage space in the computer.
When data acquisition is successfully
completed, a ``RESET'' signal is
distributed to the entire system to turn it
back to the read-enable state for
the next event.

The INTE signal instructs the DAQ system
to read out a valid event.
For DAQ efficiency and data suppression
reasons, only those FADC modules
with valid data are read out.
This is achieved when an ``EVENT'' signal is issued from an AS module
via its respective FADC module to the DAQ system.
This signifies at least one channel having
a data record above LoThr, and data
will be read out from all 16~channels in
this module. Software zero suppression
is performed online and instantaneously
to those
channels without any signals above LoThr.

\subsection{Trigger System}

\begin{figure}
\epsfig{file=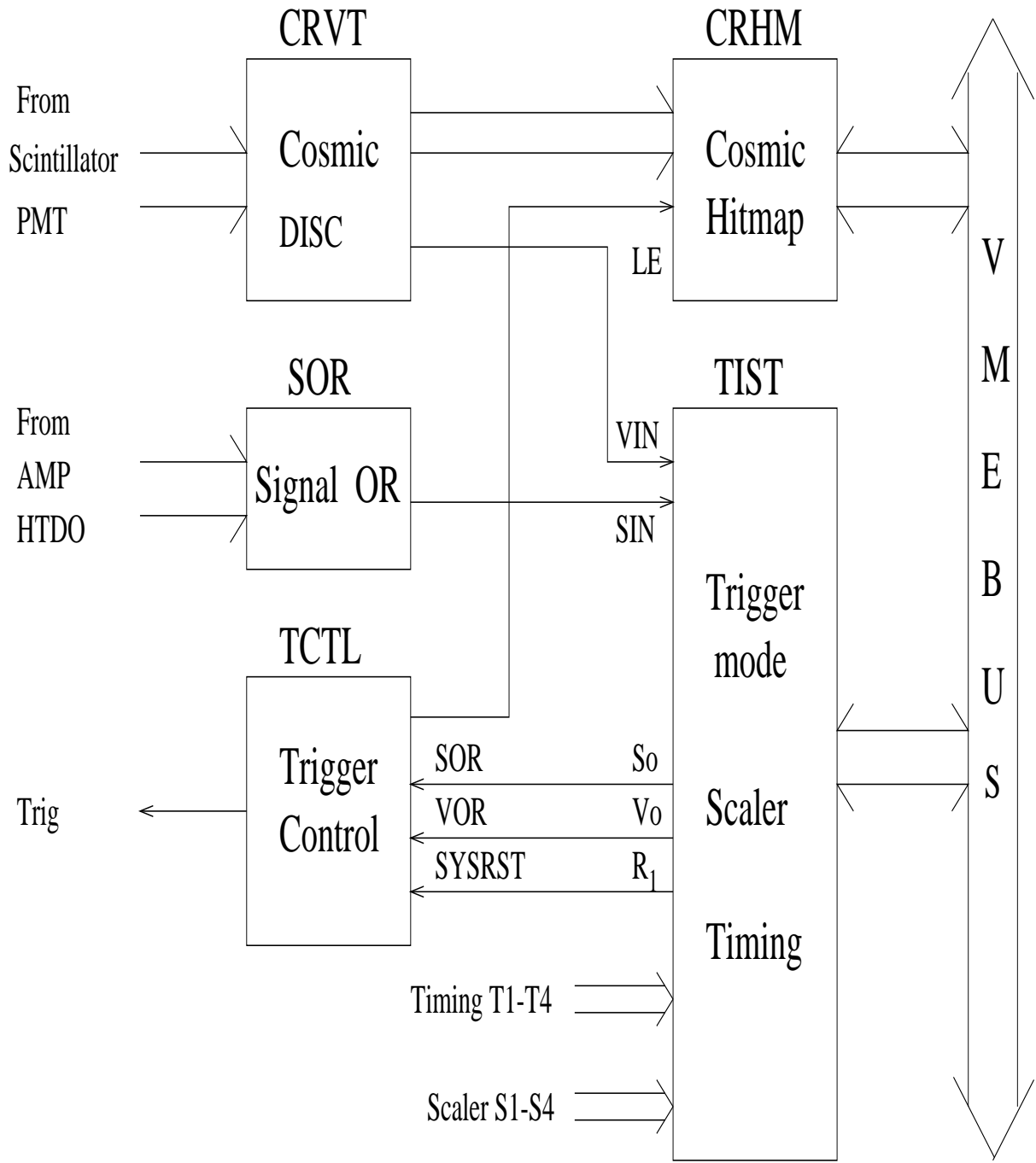,height=18cm,width=1.0\textwidth}
\vspace{0.5cm}
\caption[Trigger]{\label{fig:trigger} 
A schematic diagram  of the trigger system.}
\end{figure}

The trigger system is responsible for selecting
events to be recorded. The schematic diagram
is displayed in Figure~\ref{fig:trigger}.
A logic FAN-IN of the
signals from all the cosmic veto scintillators,
after a coincidence requirement within
individual panels, generates the ``$\rm{V_{IN}}$'' 
pulse.
The veto action is extended over a
hardware select-able ``veto period'',
typically of 100~$\mu$s, to get rid of 
background due to delayed cosmic-ray 
neutron-induced activities. 
The signal-definition  line ``$\rm{S_{IN}}$'' is
a logic FAN-in from  selected
signals from the CsI(Tl) target.

Four software select-able trigger modes
are implemented to serve various
data taking purposes: 
$\rm{ T1 = S_{IN} }$,
$\rm{ T2 = V_{IN} }$,
$\rm{ T3 = S_{IN} + V_{IN}}$, and
$\rm{ T4 = S_{IN} + \bar{V}_{IN} }$.
The ``physics trigger'' is provided by T4,
while T3 is used for dedicated data taking for
cosmic rays, while T1 and T2 are for diagnostics
and debugging purposes.
In steady-state data taking, the ``trigger menu''
consists mainly of T4 with a small sample T3 and
calibration pulser events taken once every
several hours for monitoring and calibration
purposes.

The definition of $\rm{S_{IN}}$ depends
on the selection requirements based
on the hit-pattern input from the
various CsI(Tl)+PMT channels.
For low count rate applications such as
neutrino physics experiments, the
trigger conditions can be kept loose
and minimal, and can be fine-tuned
to be optimized for specific 
physics focus.

The ``minimal bias'' trigger is 
a simple coincidence between 
both PMTs corresponding to the same
crystal module, whose pulse height
are above the HiThr discriminator level.
This is realized by having
the ``left'' PMTs connected to one AS module,
and the ``right'' ones to another.
A coincidence from 
the HDTO lines of these modules,
followed by a logic FAN-IN
circuitry provides the $\rm{S_{IN}}$ line.
Detailed selection of signal
events from the background (for instance,
events with a single hit versus those
with multiple hits)
can be carried out with subsequent offline  analysis.

The trigger system is equipped with other
diagnostic tools, like  
hit-pattern units for the cosmic-rays veto panels,
scalers for recording various count rates (like
the trigger and veto rates),
as well as TDCs for measuring various timing
(like the time between cascade sequence,
as well as that between the $\rm{S_{IN}}$
and $\rm{V_{IN}}$ lines).
The time between the most recent veto pulse 
and a valid trigger is recorded for
diagnostic purposes.


\section{Software System}
\label{sect::software}

The tasks of the software system are
to perform data acquisition
of the electronics discussed in Section~\ref{sect::ele},
to operate a slow control system  for
the high voltage supplies and other
ambient parameters, 
as well as to provide event display and monitoring 
graphic output.
These functions 
are constructed in different 
programming languages and with
different software tools, and work 
together in one single PC which runs 
on the Linux operation system. 
For security reasons, 
remote access and control to the reactor site
can only be done with a 
telephone line dial-up solution. 
 
\begin{figure}
\epsfig{file=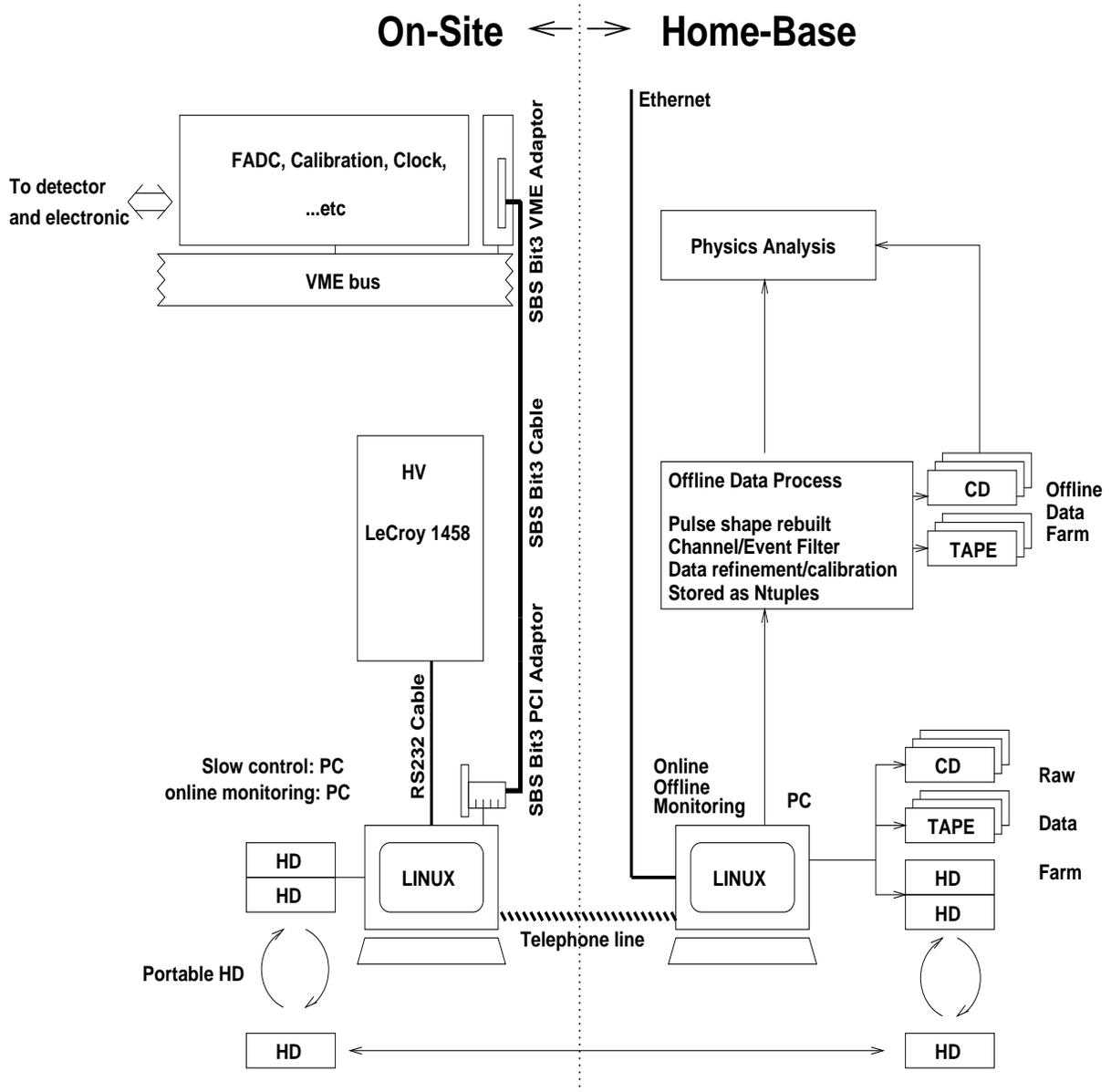,height=16cm,width=1.0\textwidth}
\vspace{0.5cm}
\caption[Calpulser]{\label{fig:ASNP} 
A schematic diagram of the online and offline
software architecture, indicating the data flow
}
\end{figure}

The schematic framework of the software
system, as well as the data flow, are depicted in 
Figure~\ref{fig:ASNP}.
The FADC data are recorded on 
hard disks. Only minimal bias selection
criteria are applied online.
Event display for data-taking quality checks   
are provided for the on-site data taking.

At steady-state data-taking, 
the experiment is expected
to be attended manually once per week.
Filled-up hard disks will be 
replaced by new ones
and brought back to the home-base laboratory
where the data are copied into 
CDs and Exabyte tapes as permanent storage media.
Meanwhile HBOOK\cite{bib:HBOOK} N-tuples 
are prepared for the 
storage of refined data which passes 
some technical event filter criteria and contains 
useful reconstructed quantities,
such as energy depositions and longitudinal positions,  
from the pulse shape information.
The refinement process of raw data to 
N-tuples can be iterated as many times as needed 
based on better
understanding on experimental 
running conditions and detector calibrations.
The calibrated N-tuples are then 
distributed to collaborating laboratories
for further physics analysis.   

Details of the various
components are described below.

\subsection{Data Acquisition}

\begin{figure}
\epsfig{file=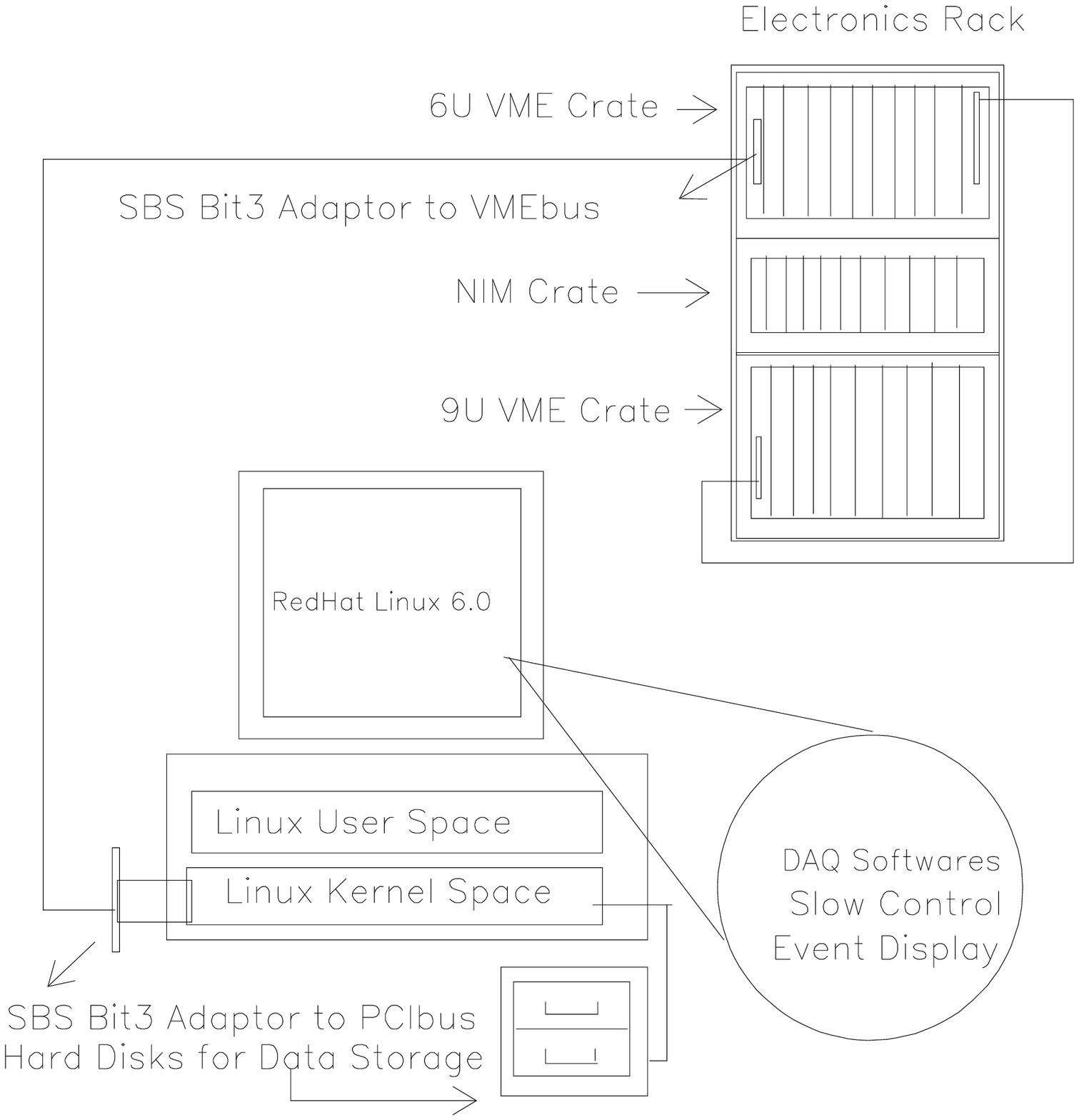,height=16cm,width=1.0\textwidth}
\vspace{0.5cm}
\caption[Calpulser]{\label{fig:SBS} 
A schematic diagram of the data acquisition system,
and the inter-connections with the electronics crates.}
\end{figure}

The schematic architecture of the 
data acquisition (DAQ) software system 
is shown in Figure~\ref{fig:SBS}.
The system 
provides control to the experimental 
running parameters, 
accesses valid data from the VME electronics modules, 
and saves them on to hard disks.

The host computer is a  PC 
running with the Pentium III-500 processor
using RedHat Linux as the Operation System.
The PC masters the VME 
slave modules on both the 6U and 9U VME crates via 
a commercial Adaptor system\footnote{SBS Bit-3 Adaptor} 
which provides communication
between the VME-bus and the PCI-bus.

The DAQ software
can be divided into two components
in kernel and user space.
The program written at the 
Linux kernel space 
provides low-level access to the VME-PCI Adaptor,
and serves as a device driver which 
uses the kernel functions and provides
the entry-point where user-space programs
can communicate with the hardware. 

The Tk/TcL-based graphics user interface 
package is equipped with
various operational dialogues,
including Start/Pause/Stop buttons and 
pull down menus for other functionalities  
for electronic module tests and regular DAQ operation.
Various input fields are provided,
where running conditions and parameters 
can be configured, file names and 
directory paths can be specified, and  operational comments 
can be recorded. 
A function for 
pre-scheduling different trigger modes is 
also available which enables automatic
selection of trigger conditions within
the same data taking period.
All control buttons in the dialogues 
are responded by their respective call-back functions 
at the kernel level of Linux.

\subsection{Slow Control}


The slow control system 
serves to record and monitor 
the high voltage (HV) power supplies for
the CsI(Tl) and veto scintillator PMTs,
as well as other ambient and operation parameters.

The high voltage system is based
on a dedicated main-frame crate\footnote{LeCroy 1458}
operating on special HV modules\footnote{LeCroy 1461}.
Each HV supply is distributed to two PMTs at
the detector level.
There are 12 HV supplies in each module
and 16 modules can be inserted into one crate,
so that a crate can handle 384 channels altogether.

The crate 
is internally equipped with a micro processor, 
adequate RAM space, and communication ports 
which can receive 
and execute
external commands or return the 
voltage and current status of the individual
channels.
The serial port communication 
are adopted (the other option is the Ethernet)
via a RS-232C cable to the DAQ PC.
In principle, one PC can host up to 
4 serial ports with each port driving one crate,
such that future expansion of the system
is straight-forward.

A VT100-based socket program is used to
communicate with the PC serial ports using
multiple Linux threads. Control and monitoring 
is achieved both locally on the console PC 
or remotely via the PPP (Point-to-Point Protocol)
dial-up connection from home-base laboratories,
as described in Section~\ref{sect::connect}.  

Slow control data such as output voltages
and currents measured by the main-frame
processor are recorded by the socket 
program on the hard disks of the DAQ PC,
at a typical frequency of once per 10~minutes. 
Subsequent monitoring software packages
can detect the voltage or current fluctuation,
and warning or fatal error messages will be 
issued in case of alarms. Appropriate
actions will be taken automatically,  like
voltage ramp-up for a tripped HV channel.

Other ambient conditions like temperature
readings from thermostats,
as well as operating parameters like
the HiThr and LoThr values, are read out
by a QDC module$^{\ref{qdc}}$.
The cumulative count rates over a data-taking
period from the CsI(Tl)
target and the veto panels (the
$\rm{S_{IN}}$ and $\rm{V_{IN}}$
lines, respectively)
are measured by a scaler module\footnote{CAEN V260}.
Both of these are read out by the main DAQ program,
at a typical frequency of once every hour.

The slow control functions
consume very little CPU time,
and therefore does not affect 
the DAQ speed, such that both operations can
both be handled by a single PC
without interference on the actual performance.

\subsection{Display and Monitoring Graphics}


A PAW-based~\cite{bib:PAW} event display and monitoring program
enable data quality check on-site from the  
information provided by the 
pulse shape, energy spectra, 
hit map, and calibration pulser analysis.
Both real-time and on-disk raw data can be accessed.  
The graphics panels paged with 
these specific functional buttons 
are implemented with PAW executable files, 
and can be linked together and switched 
with each other on the HIGZ~\cite{bib:HIGZ} window.

A clone of this program is duplicated on an 
off-site PC with an extra 
function which allows automatic data transfer 
(using a batch ftp shell script) 
from the on-site DAQ PC. To avoid the traffic 
caused by the low bandwidth of the 
networking telephone line, as discussed in  
Section~\ref{sect::connect},
only a summary data in 
the format of HBOOK N-tuples and histograms  are
transferred back to home-base laboratories
on a daily basis for offline monitoring 
and analysis.

\subsection{\label{sect::connect}Network Connection}

Owing to security reasons,
the on-site computer can only 
be accessed via telephone line from home-based
laboratories, as depicted in Figure~\ref{fig:ASNP},
to decouple the connections from 
the reactor plant's
internal Ethernet network. 
The PPP system is adopted for the network connection 
of the on-site 
DAQ PC and the off-site PC, which serves the 
entry point for other off-site computers. 
Under the PPP client-server mode, the on-site 
PC is configured as the PPP server, whose modem 
can automatically pick up a dial-up connection 
request from the off-site PPP client, which 
get its IP address from the server.

Both the server and client are equipped 
with 56k bit-per-second(bps) modems, and the bandwidth is 
limited by the quality (9.6k~bps) of the existing 
internal telephone circuitry of of the reactor
plant. 
An average networking 
speed is around 4.8k~bps is achieved.
This is adequate for monitoring purposes
when no large amount of data transfer
is involved.

\section{Performance}

The electronics and data acquisition
systems with a 
total of 200 readout channels
has been in robust operating conditions 
taking data from the prototype crystal modules
at the laboratory.
The readout scheme essentially preserves
all energy and timing information recorded
by the detector, providing powerful
diagnostic tools for rare-search type 
experiments.

\begin{figure}
\centerline{
\epsfig{file=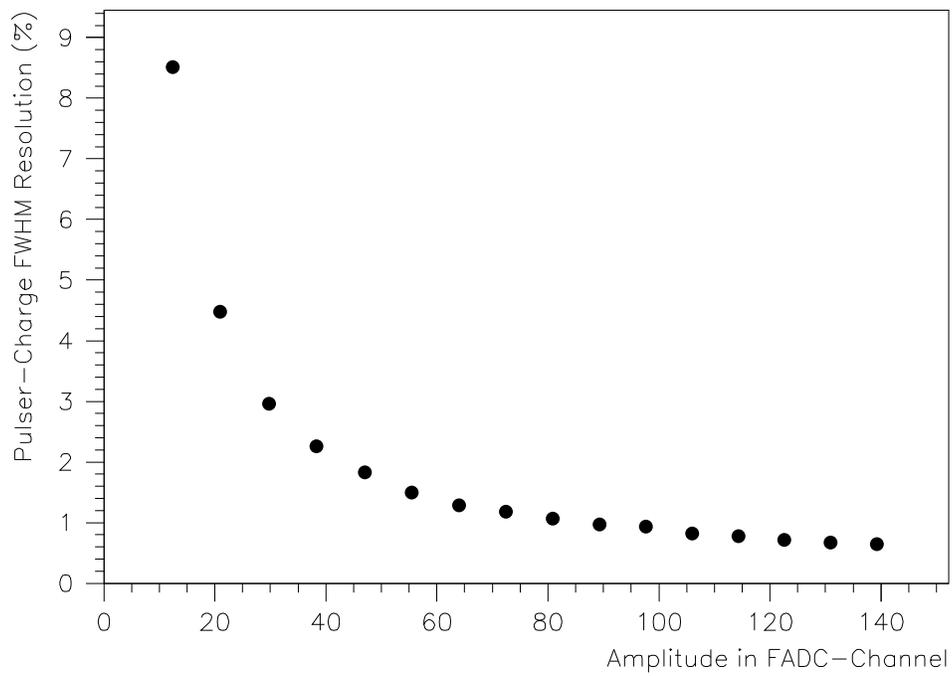,width=15cm}
}
\caption{
The variation of FWHM resolution of
the calibration pulser-charge as
a function of pulse amplitude in
units of FADC-channel. The full range of
255 corresponds to 2~V at the input level
of the FADC.
}
\label{fig:pulserreso}
\end{figure}

Using the input from the precision
calibration pulser, the overall linearity of
the response of the electronics
and data acquisition systems
is better than 0.3\%. The
RMS of the pedestal is equivalent
to 0.30~FADC-channel corresponding
to 2.3~mV baseline noise
at the input level of the
FADC. The resolution of the 
measured pulser-charge as a
function of pulse amplitude, in the unit of FADC-channel,
is displayed in Figure~\ref{fig:pulserreso}.
The readout capabilities
are much better than the 
detector response such that
the overall performance
parameters of the experiment
are limited by the detector
hardware.

\begin{figure}
\centerline{
\epsfig{file=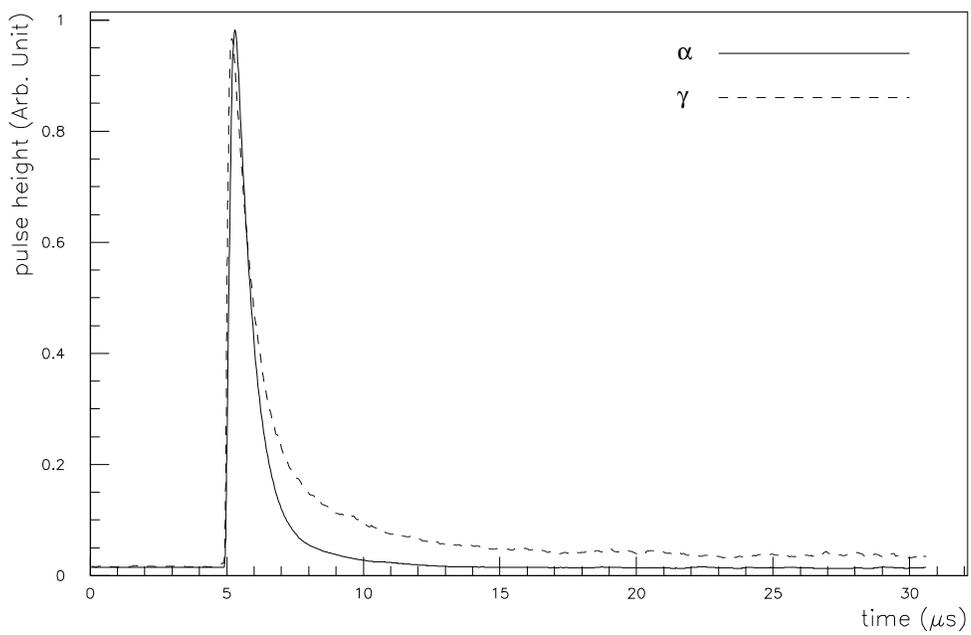,width=15cm}
}
\caption{
The average pulse shape
of events due to $\gamma$-rays and $\alpha$-particles
as recorded by the
FADC module. Their different decay times provide
pulse shape discrimination capabilities.
}
\label{fig:psd}
\end{figure}

\begin{figure}
\centerline{
\epsfig{file=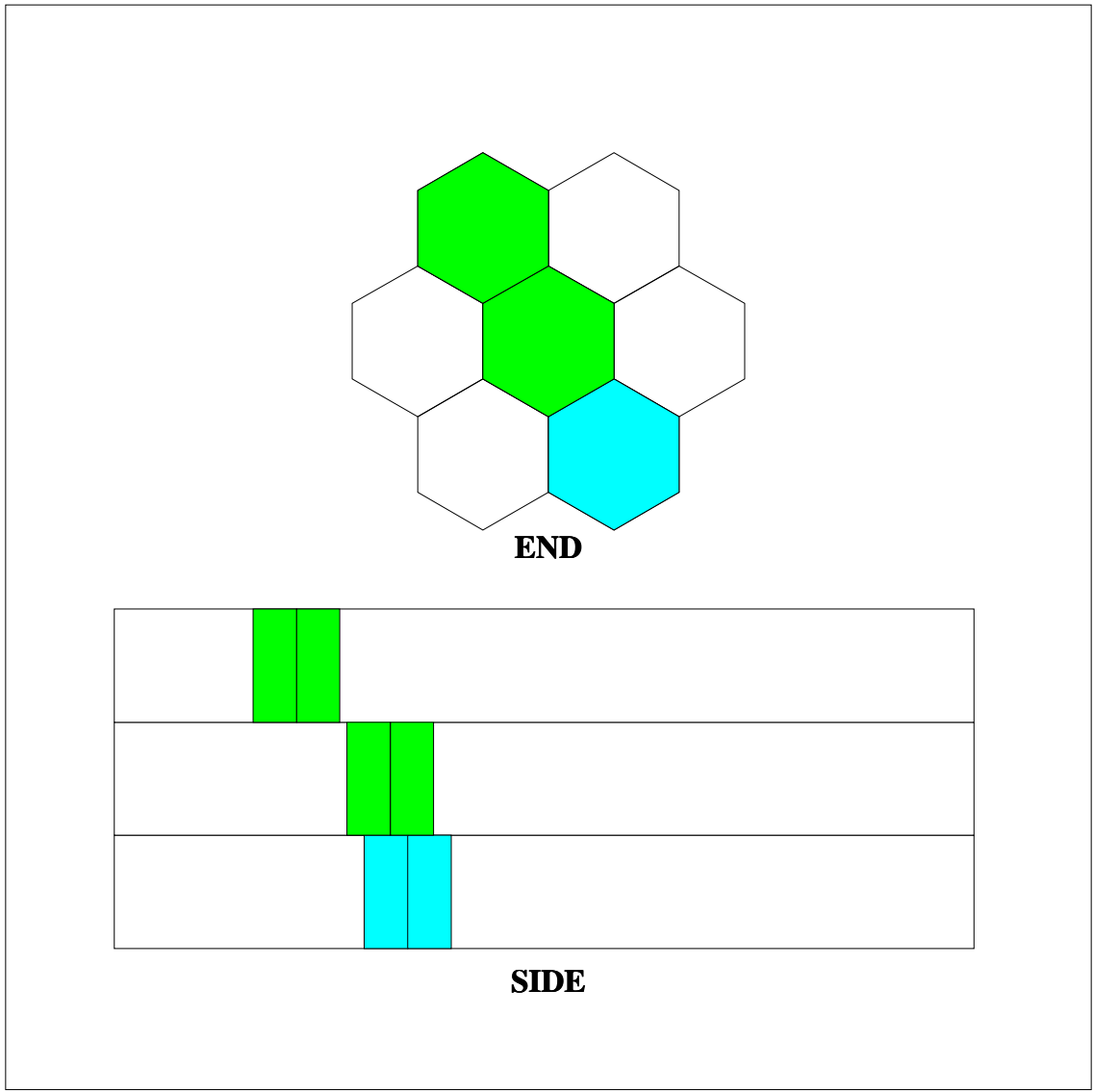,width=15cm}
}
\caption{
A typical cosmic-muon event recorded
by a 7-module/14-channel  prototype
detector, showing both the end and side
views.
}
\label{fig:muon}
\end{figure}

The typical signals
from the CsI(Tl)+PMT 
and the FADC outputs 
are displayed in Figure~\ref{fig:signals}a 
and \ref{fig:signals}b, respectively.
The fall times are different between $\gamma$/e 
and $\alpha$ events, as shown in
Figure~\ref{fig:psd}, providing a
basis of particle identification.
A typical cosmic-ray muon traversing
a prototype detector set-up with
7 crystal modules is presented in Figure~\ref{fig:muon}.

\begin{figure}
\centerline{
\epsfig{file=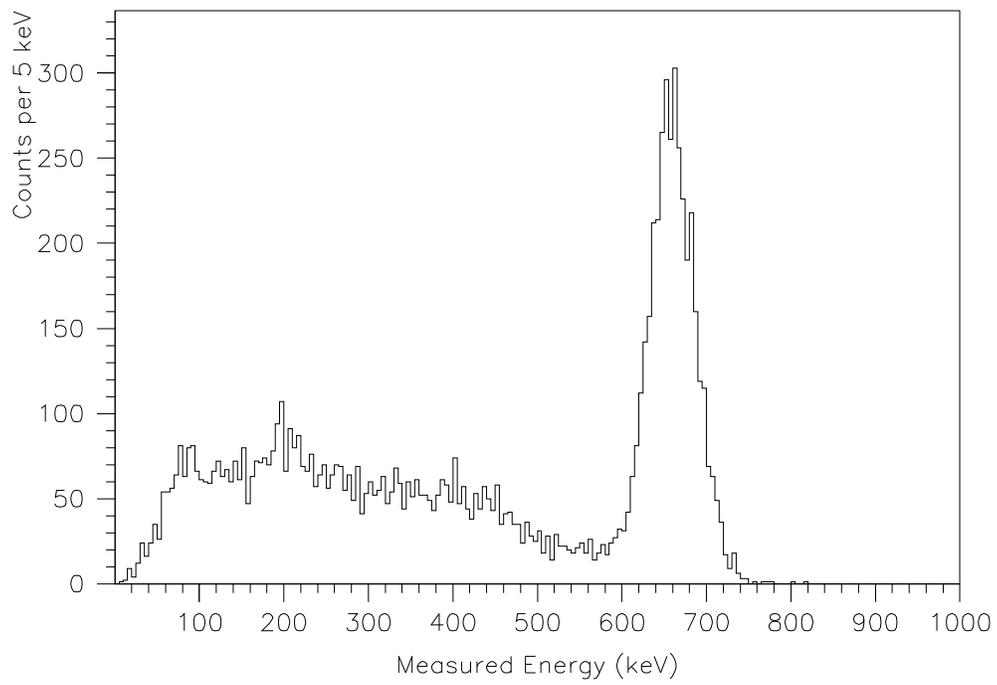,width=15cm}
}
\caption{
The measured energy spectrum
of a $^{137}$Cs source.
The FWHM resolution at 660~keV is about 10\%.
}
\label{fig:cs137}
\end{figure}

The integrated sum of the signals
from both ends of a crystal module gives the energy
information of the event. 
A typical
energy spectrum due to
a $^{137}$Cs source is depicted in Figure~\ref{fig:cs137}.
A energy threshold of less than 50~keV is achieved.
The full-width-half-max (FWHM) 
resolution at 660~keV is about 10\%.
This energy 
corresponds to pulses with around 50~FADC-channels
in amplitude, which implies, based
on Figure~\ref{fig:pulserreso},
a readout
contribution of 1.6\% to the resolution effects
at the $^{137}$Cs energy.

The system dead time is due to the contributions
of two factors: (1) veto dead time, which is
the veto rate multiplied by the
inactive time per event (typically 100~$\mu$s),
and (2) DAQ dead time, which is
the time needed to read out the FADC and other modules
(that is, the time after the STROB gate till the RESET pulse,
as displayed in Figure~\ref{fig:Timing}).

The DAQ dead time 
depends on the complexity of the events.
To illustrate the range,
the typical dead time is 16~ms to read
out and zero-suppress the simplest
event where there is a
single hit without delayed cascade,
requiring only one FADC per event to be handled.
The corresponding dead time for
the read-out
of 9 FADCs each having 7 delayed cascade per event
is 700-800~ms.
Typically, in situations like
calibration data taking with 
radioactive sources which give
high count rate but simple hit-pattern,
and where
the delayed cascade functions are disabled,
a DAQ 
rate of 100~Hz can be sustained.


\section{Summary}


An  electronics and data acquisition system
for low energy neutrino physics experiment,
based on CsI(Tl) scintillating crystal 
as detectors with 100$-$500 readout channels,
has been successfully designed, built and
commissioned. It is modular by design
and flexible in applications, such that
the system can be
adapted, as a whole or in parts, for other purposes.

There are several possible future upgrades 
which can further
enhance its capabilities.
The amplifier/shaper modules are equipped 
with shift registers such that a hit-pattern
can be extracted if necessary. The readout
time can then be reduced with this prior
knowledge of what channels to read.
Similarly, the trigger conditions
employed for the current experiment are
very loose and with minimal bias. 
With the exact hit-pattern information extracted,
more sophisticated trigger selection
criteria can be devised. 
Both functions are not critical for low count
rate, low occupancy experiments, but 
may be desirable for higher trigger rate
applications like in a high energy physics
environment.

The current application requires a 
large dynamic
range from about 10~keV (several photo-electrons)
to 50~MeV  (cosmic-ray muons).
Although the FADC utilizes an
8-bit digitization scheme,
its capabilities are beyond a $2^8$=256 fold
dynamic range, since the entire pulse with
a few hundred data-points are measured for
every event.
Over-scale signal manifest itself as event with
a ``flat-top'', the duration of which provides
a measurement to the energy of the event. 
Using this feature~\cite{bib:dynrange},
the desired range can be covered even with
a readout system with an 8-bit resolution
in amplitude.

     
The authors are grateful to the technical staff of our
institutes for invaluable support.
This work was supported by contracts
NSC~88-2112-M-001-007,
NSC~89-2112-M-001-028 and
NSC~89-2112-M-001-056
from the National Science Council, Taiwan,
as well as
19975050 from the
National Science Foundation, China.

\clearpage

\clearpage

\clearpage

\clearpage

\clearpage

\clearpage

\clearpage

\clearpage

\clearpage

\clearpage

\clearpage

\clearpage

\clearpage

\clearpage

\end{document}